\documentclass[12pt]{article}

\usepackage{graphics,amssymb,epsfig,float}
\usepackage[usenames,dvips]{color}
\usepackage{graphicx}
\usepackage{epsfig}
\usepackage{rotating}
\usepackage{dcolumn}
\usepackage{bm}
\usepackage{cite}
\usepackage{amsmath}
\usepackage{array,hhline}

\def\lsim{\;\raise0.3ex\hbox{$<$\kern-0.75em\raise-1.1ex\hbox{$\sim$}}\;}
\def\gsim{\;\raise0.3ex\hbox{$>$\kern-0.75em\raise-1.1ex\hbox{$\sim$}}\;}
\def\beq{\begin{equation}}   \def\eeq{\end{equation}}
\def\ba{\begin{array}}       \def\ea{\end{array}}
\def\bea{\begin{eqnarray}}   \def\eea{\end{eqnarray}}

\def\nl{\newline}
\def\k{\kappa}
\def\l{\lambda}

\def\noi{\noindent}

\begin{document}

\begin{titlepage}
\begin{flushright}
LPT Orsay 11-65 \\ 
\end{flushright}

\begin{center}
\vspace{1cm}
{\Large\bf Higgs Bosons in the Next-to-Minimal Supersymmetric Standard
Model at the LHC} \\
\vspace{1cm}

{Ulrich Ellwanger$^1$}\\
\vspace*{.5cm}
$^1$  Laboratoire de Physique Th\'eorique, UMR 8627, CNRS and
Universit\'e de Paris--Sud,\\
B\^at. 210, F-91405 Orsay, France \\

\end{center}

\vspace{1cm}

\begin{abstract}
We review possible properties of Higgs bosons in the NMSSM, which allow
to discriminate this model from the MSSM: masses of mostly
Standard-Model-like Higgs bosons at or above 140~GeV, or enhanced
branching fractions into two photons, or Higgs-to-Higgs decays. In the
case of a Standard-Model-like Higgs boson above 140~GeV, it is
necessarily accompagnied by a lighter state with a large gauge singlet
component. Examples for such scenarios are presented. Available studies
on Higgs-to-Higgs decays are discussed according to the various Higgs
production modes, light Higgs masses and decay channels.
\end{abstract}

\end{titlepage}

\section{Introduction}

One of the main goals of the Large Hadron Collider (LHC) is the
detection of the Higgs boson, or of at least one of several Higgs bosons
if corresponding extensions of the Standard Model (SM) are realized in
nature. These searches depend crucially on the Higgs masses, production
cross sections and the Higgs decays.

In the case of the SM, the production cross sections and decay branching
ratios are quite well known as functions of the unknown Higgs mass
\cite{Djouadi:2005gi}. In the Minimal Supersymmetric Standard Model
(MSSM) with its extended Higgs sector and parameter space, these
quantities have been studied as well and it seems that at least one of
the Higgs bosons cannot be missed at the LHC
\cite{Ball:2007zza,Aad:2009wy,Djouadi:2005gj} (we speak of a so-called
no-lose theorem). There exist, however, well motivated scenarios with
somewhat more extended Higgs sectors, as the Next-to-Minimal
Supersymmetric Standard Model (NMSSM, see
\cite{Maniatis:2009re,Ellwanger:2009dp,Teixeira:2011cz} for recent
reviews), where the Higgs production rates and decays can differ
strongly from both the SM and the MSSM. It is very important to be aware
of the possibility of such unconventional properties of Higgs bosons;
otherwise the absence of a signal in standard Higgs search channels, or
unusual signals, may be completely misinterpreted.

Typical such unconventional properties of Higgs bosons in the NMSSM are
Higgs-to-Higgs decays, Higgs boson with reduced couplings to gauge
bosons, and/or Higgs masses incompatible with the MSSM. In the last
years, many studies have been performed in order to investigate which
scenarios are possible in the NMSSM, and by means of which signals they
could be detected. Note that the interest in such studies is twofold:
In some cases, it can be very challenging to detect a signal of any of
the Higgs bosons of the NMSSM. In other cases a single signal
is nearly as easy to see as in the MSSM, but only a detailed study of
the complete visible Higgs spectrum can possibly allow to distinguish
the NMSSM from the MSSM: For instance, the mass of the dominantly
SM-like Higgs boson (with the largest couplings to electroweak gauge
bosons) can be larger than 140~GeV in the NMSSM, with slightly
reduced couplings to electroweak gauge bosons\footnote{Such scenarios
can be consistent with recent results reported by the CMS and ATLAS\nl
collabo\-rations \cite{cms_higgs,atlas_higgs}.}.

In the present paper we discuss the status of such NMSSM-specific Higgs
properties and searches. We review the various possible scenarios, and
the available studies on corresponding search strategies for Higgs
bosons.

The Higgs sector of the NMSSM consists of two SU(2) doublets $H_u$ and
$H_d$ (where, as in the MSSM, $H_u$ couples to up-type quarks and $H_d$
to down-type quarks and leptons), and one additional gauge singlet $S$.
Due to its coupling $\l S H_u H_d$ in the superpotential, a vacuum
expectation value (vev) $s$ of $S$ generates a supersymmetric mass term
$\mu_\mathrm{eff}=\l s$ for $H_u$ and $H_d$. Since $s$ and hence
$\mu_\mathrm{eff}$ are naturally of the order of the soft Susy breaking
terms $\sim M_\mathrm{Susy}$, this solves the so-called $\mu$-problem of
the MSSM \cite{Kim:1983dt}. Furthermore, in its simplest $Z_3$ invariant
version, the superpotential of the NMSSM is scale invariant; it is in
fact the simplest phenomenologically acceptable supersymmetric extension
of the SM with this property. The NMSSM shares with the MSSM the
unification of the running gauge coupling constants at a Grand
Unification (GUT) scale, and the natural presence of a dark matter
candidate in the form of a stable lightest supersymmetric particle
(LSP).

The physical neutral Higgs sector in the NMSSM consists of 3 CP-even and
2 CP-odd states. (Here we do not consider the possibility of CP
violation in the Higgs sector.) In general, these states are mixtures of
the corresponding CP-even or CP-odd components of the weak eigenstates
$H_u$, $H_d$ and $S$, without the CP-odd Goldstone boson swallowed by
the massive $Z$ boson.

Many analyses of the Higgs sector of the NMSSM 
\cite{Ellis:1988er,Drees:1988fc,Ellwanger:1993xa,Kamoshita:1994iv,
Ellwanger:1995ru,Franke:1995xn,King:1995ys,Gunion:1996fb,Ham:1996sf,
Krasnikov:1997nh,Ellwanger:1999ji,Ellwanger:2001iw,Ellwanger:2003jt,
Ellwanger:2005uu,Moretti:2006sv,Dermisek:2007ah} pointed out that the
physical eigenstates in the CP-even sector can well be strong mixtures
of SU(2) doublet and singlet states with reduced couplings to gauge
bosons. This motivated studies on the detectability of one or many Higgs
states at LEP \cite{Ellwanger:1995ru, Franke:1995xn,King:1995ys,
Gunion:1996fb,Ham:1996sf}, the LHC \cite{Gunion:1996fb,
Krasnikov:1997nh, Ellwanger:2001iw,Ellwanger:2003jt,Ellwanger:2005uu,
Moretti:2006sv} and a more energetic linear collider
\cite{Kamoshita:1994iv} (see also \cite{Gunion:2003fd} for searches for
Higgs bosons beyond the SM at linear colliders).

As mentioned first in \cite{Gunion:1996fb}, Higgs-to-Higgs decays can be
important for Higgs detection in certain regions of the parameter space
of the NMSSM. Notably the lightest CP-odd state $A_1$ can play the role
of a pseudo-Goldstone boson \cite{Dobrescu:2000jt,Dobrescu:2000yn} whose
small mass can lead to dominant $H \to A_1 A_1$ decays of CP-even states
$H$ \cite{Dermisek:2005ar,Dermisek:2005gg,Dermisek:2006wr}. The
possibility of Higgs-to-Higgs decays inhibited to establish an
all-embracing no-lose theorem for the NMSSM
\cite{Ellwanger:2001iw,Ellwanger:2003jt}, and triggered numerous studies
on Higgs detection in such circumstances. The proposed strategies depend
on the masses and branching ratios of the involved Higgs bosons, and
will be reviewed in Chapter~4.

A list of benchmark points corresponding to unconventional scenarios in
the Higgs sector of the NMSSM was proposed in \cite{Djouadi:2008uw},
where the then available search strategies have been summarized.
Non-standard Higgs boson decays were also reviewed in
\cite{Chang:2005ht,Chang:2008cw,Dermisek:2010tg} and, in the light of
the fine-tuning problem, in \cite{Dermisek:2009si}. Both singlet-doublet
mixings and Higgs-to-Higgs decays allow for SM-like CP-even Higgs bosons
with masses well below 114~GeV, compatible with LEP constraints
\cite{Schael:2006cr} alleviating the ``little fine-tuning problem'' of
the MSSM \cite{Dermisek:2005ar,Dermisek:2007yt,Dermisek:2007ah,
Dermisek:2009si,Ellwanger:2011mu}. The reduced Higgs couplings, Higgs
production rates at the LHC and Higgs branching ratios in the NMSSM have
recently been studied in \cite{Mahmoudi:2010xp}.

In the next Chapter we briefly review the Higgs sector of the NMSSM, and
present constraints. Chapter~3 is dedicated to NMSSM scenarios which
allow to detect a Higgs boson in ``standard'' Higgs search channels. We
focus on scenarios where signals can be visible, but where the
properties of the Higgs states (masses above $\sim 140$~GeV, branching
ratios or the number of distinct states) allow potentially to
distinguish the NMSSM from the MSSM. These scenarios occur typically in
the case of singlet/doublet mixings in the Higgs sector of the NMSSM. In
Chapter~4 we consider scenarios which require dedicated search
strategies, notably in the cases of dominant Higgs-to-Higgs decays.
Conclusions are given in Chapter~5.

\section{The Higgs sector of the NMSSM}

The NMSSM differs from the MSSM due to the presence of the gauge singlet
superfield $S$. In the simplest $Z_3$ invariant realisation of the
NMSSM, the Higgs mass term $\mu H_u H_d$ in the superpotential
$W_{MSSM}$ of the MSSM is replaced by the coupling $\lambda$ of $S$ to
$H_u$ and $H_d$ and a self-coupling $\kappa S^3$.  Hence, in this
simplest version the superpotential $W_{NMSSM}$ is scale invariant, and
given by:
\beq\label{eq:1}
W_{NMSSM} = \lambda \hat S \hat H_u\cdot \hat H_d + \frac{\kappa}{3} 
\hat S^3 + \dots\; ,
\eeq
where hatted letters denote superfields, and the dots denote the
MSSM-like Yukawa couplings of $\hat H_u$ and $\hat H_d$ to the quark and
lepton superfields. Once the real scalar component of $\hat S$ develops
a vev $s$, the first term in $W_{NMSSM}$ generates an effective
$\mu$-term
\beq\label{eq:2}
\mu_\mathrm{eff}=\lambda s\; .
\eeq
The phenomenological constraint $\mu_\mathrm{eff} \gsim 100$~GeV from
the non-observation of charginos implies $s \gsim 100\ \mathrm{GeV}/\l$.

The soft Susy breaking terms consist of mass terms for the Higgs bosons
$H_u$, $H_d$ and $S$, and trilinear interactions (omitting squarks and
sleptons)
\beq\label{eq:3}
 -{\cal L}_\mathrm{Soft} =
m_{H_u}^2 | H_u |^2 + m_{H_d}^2 | H_d |^2 + 
m_{S}^2 | S |^2 +\Bigl( \lambda A_\lambda\, H_u
\cdot H_d \,S +  \frac{1}{3} \kappa  A_\kappa\,  S^3 \Bigl)+
\mathrm{h.c.}\; .
\eeq
Expressions for the mass matrices of the physical CP-even and CP-odd
Higgs states -- after $H_u$, $H_d$ and $S$ have assumed vevs $v_u$,
$v_d$ and $s$ and including the
dominant radiative corrections -- can be found in
\cite{Ellwanger:2009dp} in will not be repeated here; below we just
recall some relevant properties of the physical states. (We will use
$\tan\beta = v_u/v_d$ and $v^2=v_u^2 + v_d^2 \simeq (174\
\mathrm{GeV})^2$.)

In the CP-even sector we find three states which are mixtures of the
real components of $H_u$, $H_d$ and $S$. The state $h$ with the largest
(often nearly SM-like) coupling to the electroweak gauge bosons has a
mass squared $M_h^2$ given by\footnote{We use $h$ for the mostly SM-like
CP-even Higgs boson, but $H$ or $H_i$ for general CP-even Higgs bosons.}
\beq\label{eq:4}
M_h^2 = M_Z^2\cos^2 2\beta + \l^2v^2\sin^2 2\beta + \mathrm{rad.\
corrs.} + \Delta_\mathrm{mix}\; ,
\eeq
whereas the diagonal matrix element in the singlet sector is given by
(assuming $s \gg v_u,\,v_d$)
\beq\label{eq:5}
M_{SS}^2 \simeq \k s(A_\k + 4\k s)\; .
\eeq
The term $\Delta_\mathrm{mix}$ in (\ref{eq:4}) originates from
singlet-doublet mixing and becomes for weak mixing
\beq\label{eq:6}
\Delta_\mathrm{mix} \simeq \frac{4\l^2 s^2 v^2 (\l - \k \sin 2\beta)^2}
{\overline{M}_h^2 - M_{SS}^2}
\eeq
where $\overline{M}_h^2$ is given by $M_h^2$ without the mixing term.
Several remarks are in order.

First, neglecting singlet-doublet mixing, $M_h$ can be larger than in
the MSSM due to the second term in (\ref{eq:4}): up to $\sim 140$~GeV
\cite{Ellwanger:2006rm} if the running coupling $\l$ is assumed to
remain perturbative below the GUT scale, but up to $\sim 300$~GeV
\cite{Barbieri:2006bg} if this assumption is given up.

Second, depending on the unknown parameters as $A_\k$ and $\k s$,
$M_{SS}^2$ -- and hence the mass of the singlet-like CP-even state --
can be larger or smaller than $\overline{M}_h^2$. For $M_{SS}^2 >
\overline{M}_h^2$ we have $\Delta_\mathrm{mix} < 0$ in (\ref{eq:4}).
Hence it is not guaranteed that the contribution to $M_h^2$ from the
NMSSM specific terms in (\ref{eq:4}) -- the sum of the second and forth
terms on the right hand side -- is positive. (Of course, the negative
contribution from $\Delta_\mathrm{mix}$ vanishes if, accidentially, $\l
\sim \k \sin 2\beta$.)

For $M_{SS}^2 < \overline{M}_h^2$, $h$ is actually the second lightest
Higgs state, and the mass of the singlet-like CP-even state is typically
below 114~GeV. Now we have $\Delta_\mathrm{mix} > 0$ in (\ref{eq:4}),
which can augment the mass $M_h$ well above 140~GeV even for larger
$\tan\beta$, where the second NMSSM specific term in (\ref{eq:4})
becomes small.

Now the mixing angle (i.e. the coupling to the $Z$ boson) of the
singlet-like CP-even Higgs state is constrained by LEP
\cite{Schael:2006cr}: The non-observation of a signal at LEP leads to
upper bounds on $\xi^2 \equiv \bar{g}^2 \times \overline{BR}(H \to b
\bar{b})$ as function of $M_H$, where $\bar{g}$ is the reduced coupling
of $H$ to $Z$ (normalized with respect to the SM), and $\overline{BR}(H
\to b \bar{b})$ the branching ratio into $b \bar{b}$ normalized with
respect to the SM. (The singlet-like CP-even state will still have
$\overline{BR}(H \to b \bar{b}) \sim 1$.) For convenience we have
reproduced the corresponding figure from \cite{Schael:2006cr} as
Fig.~\ref{fig:1} below.

\begin{figure}[ht!]
\begin{center}
\includegraphics[scale=0.5]{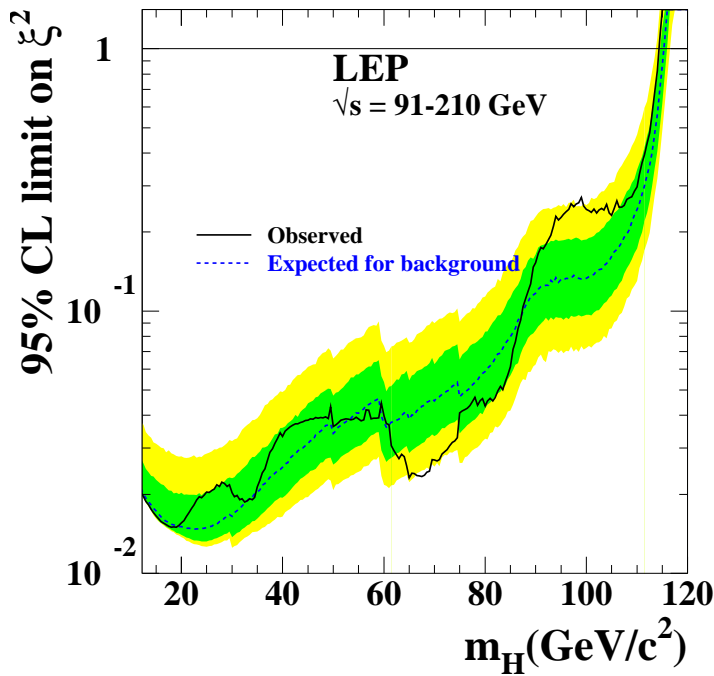}
\end{center}
\caption{Upper bounds on $\xi^2 \equiv \bar{g}^2 \times \overline{BR}(H
\to b \bar{b})$ from LEP \cite{Schael:2006cr}. Full line: observed
limit; dashed line: expected limit; dark (green) shaded band: within
68\% probability; light (yellow) band: within 95\% probability.}
\label{fig:1}
\end{figure}

One can note that, due to a slight excess of events, the upper bound on
$\xi^2$ is particularly weak for $M_H$ around $95 - 100$~GeV. This
behaviour does certainly not require the presence of a Higgs state with
a corresponding mass and $\xi^2 \sim 0.25$, but it could be explained by
a such a dominantly singlet-like CP-even Higgs state (not within the
SM!).

Finally the singlet-like CP-even Higgs state can be lighter than $M_h/2$
such that the nearly SM-like Higgs $h$ could decay dominantly into a
pair of singlet-like CP-even Higgs states.

A third CP-even Higgs state has usually a mass close to the mass of one
of the CP-odd (and the charged) Higgs states as in the MSSM. Defining
$B_\mathrm{eff}= A_\l + \k s$, the CP-odd mass squared matrix has a
diagonal element
\beq\label{eq:7}
M_{P,AA}^2 = \frac{2\mu_\mathrm{eff} B_\mathrm{eff}}{\sin 2\beta}\; .
\eeq
The diagonal element of the CP-odd mass squared matrix in the singlet
sector is given by (assuming again $s \gg v_u,\,v_d$)
\beq\label{eq:8}
M_{P,SS}^2 \simeq -3\k A_\k s\; .
\eeq
Hence, depending on the unknown parameters, one can find a light
CP-odd dominantly singlet-like state in the NMSSM. Considering the full
CP-odd mass matrix, one obtains a massless physical Goldstone boson
either in the Peccei-Quinn symmetry limit $\k \to 0$, or in the
$R$-symmetry limit $A_\l,\ A_\k \to 0$
\cite{Dobrescu:2000jt,Dobrescu:2000yn,Miller:2003ay}. Hence a light
CP-odd Higgs state $A_1$ playing the role of a pseudo-Goldstone boson is
natural in the NMSSM, if any of these symmetries is approximately
realized.

Phenomenological constraints on the mass $M_{A_1}$ depend heavily on the
coupling of $A_1$ to $b$-quarks. Normalized with respect to the
corresponding coupling of the SM Higgs boson, the $A_1$-$b$-$\bar{b}$
coupling~is
\beq\label{eq:9}
X_d = \tan\beta \cos\theta_A
\eeq
where $\cos\theta_A$ denotes the SU(2) doublet component of $A_1$. 

In any of the symmetry limits one has $\cos\theta_A \sim 1/\tan\beta$
leading to $X_d \sim v/s$ (Peccei-Quinn symmetry limit) or $X_d \sim
2v/s$ ($R$-symmetry limit) \cite{Ellwanger:2009dp}, hence typically to
$X_d \ll 1$. On the other hand, if $A_1$ is coincidentally light outside
a symmetry limit, $X_d > 1$ is possible as well.

For $M_{A_1} \lsim 9$~GeV, upper bounds on $X_d$ result from the
non-observation of $\Upsilon$ decays into $A_1$ implying $X_d \lsim 1$
\cite{Dermisek:2010mg,Domingo:2010am}. Constraints on $M_{A_1}$/$X_d$
from $B$-physics are model dependend as they depend strongly on the
flavour changing $A_1$-$b$-$\bar{s}$ vertex induced by loops of
supersymmetric particles (sparticles) and on the charged Higgs mass
\cite{Domingo:2007dx}.

For $9.2\ \mathrm{GeV} \lsim M_{A_1} \lsim 10.5$~GeV, $A_1-\eta_b$
mixings become relevant \cite{Domingo:2008rr,Fullana:2007uq} with
potentially desirable implications on $\eta_b$ spectroscopy
\cite{Domingo:2009tb}. These allow to deduce (weaker) upper bounds on
$X_d$ from the non-observation of $\Upsilon$ decays
\cite{Domingo:2010am} in this range of $M_{A_1}$, but also affect the
$A_1$ decay channels which are now "inherited" to a large extend from
the $\eta_b$ decays into two gluons \cite{Domingo:2011rn}.

Clearly, for $M_{A_1} < M_h/2$, decays of the SM-like CP-even Higgs
state $h$ into $A_1\,A_1$ are relevant. The non-observation of such
decays at LEP implies lower bounds on $M_h$ depending on $M_{A_1}$:

For $M_{A_1} \gsim 10.5$~GeV, $A_1$ would decay dominantly into $b
\bar{b}$. $h \to A_1 A_1 \to 4\,b$ decays have been searched for by the
OPAL and DELPHI groups \cite{Abbiendi:2004ww,Abdallah:2004wy}
(summarized in \cite{Schael:2006cr}) leading to $M_h \gsim 110$~GeV for
$10.5\ \mathrm{GeV} \lsim M_{A_1} \lsim 55$~GeV.

For $M_{A_1} \lsim 9.2$~GeV, $A_1$ would decay dominantly into $\tau^+
\tau^-$ and is hardly affected by $A_1-\eta_b$ mixings. The decay $h \to
A_1 A_1 \to 4\,\tau$ has recently been re-analized by the ALEPH group
\cite{Schael:2010aw} implying $M_h \gsim 107$~GeV if $\xi^2 \equiv
\frac{\sigma(e^+ e^- \to Zh)}{\sigma_\mathrm{SM}(e^+ e^- \to Zh)} \times
BR(h \to A_1 A_1) \times BR(A_1\to \tau^+\, \tau^-)^2 \sim~1$, or $M_h
\gsim 100$~GeV if $\xi^2 \sim~0.5$.

Due to the mostly gluonic decays of $A_1$ in the window
$9.2\ \mathrm{GeV} \lsim M_{A_1} \lsim 10.5$~GeV\cite{Domingo:2011rn},
constraints on $M_h$ for $M_{A_1}$ in this window result mainly from the
remaining (parameter dependend) branching ratio for the ``standard''
decay $h \to b\,\bar{b}$ \cite{Schael:2006cr}, but not from limits on
final states from $h \to A_1 A_1$.

Finally the charged Higgs boson mass is given by
\beq\label{eq:10}
M_\pm^2 = M_{P,AA}^2 + v^2\left(\frac{g_2^2}{2}-\l^2\right)
\eeq
with $M_{P,AA}^2$ as in (\ref{eq:7}). Due to the last term $\sim \l^2$,
the charged Higgs boson, compared to the corresponding CP-even and
CP-odd Higgs bosons, can be somewhat lighter as in the MSSM. Lower
bounds on $M_\pm$ result from the non-observation of charged Higgs
bosons in top quark decays at the Tevatron
\cite{Aaltonen:2009ke,:2009zh} and depend on $\tan\beta$. Stronger lower
bounds on $M_\pm$ result from $B$-physics like $b \to s\,\gamma$, unless
cancellations with sparticle-induced loop diagrams occur
\cite{Domingo:2007dx}.

The full parameter space of the NMSSM includes also the decoupling limit
$\l,\ \k \to 0$. Then the vev $s$ becomes $s \sim M_\mathrm{Susy}/\k$,
where $M_\mathrm{Susy}$ denotes the order of the soft Susy breaking
terms. Hence we find $\mu_\mathrm{eff} \sim \l/\k \cdot
M_\mathrm{Susy}$, and the $\mu$-problem is still solved for $\l \sim
\k$. In this limit the singlet-like CP-even and CP-odd Higgs states
decouple and become unobservable, independent from their masses (which
remain of ${\cal O}(M_\mathrm{Susy})$). Then the NMSSM could be
distinguished from the MSSM only if the singlino-like neutralino is the
LSP, appearing as final state in all sparticle decay cascades
\cite{Ellwanger:1997jj}.

To summarize this Chapter, the following NMSSM specific scenarios are
possible in the Higgs sector:

\begin{itemize}
\item CP-even Higgs bosons: Due to possibly large singlet-doublet mixing
angles, more -- potentially three! -- CP-even states than in the MSSM
could be observable, but with reduced signal rates for any of them. The
dominantly SM-like state $h$ can be heavier than in the MSSM. A
dominantly singlet-like state with a mass below 110~GeV is compatible
with LEP constraints, and can shift upwards (due to mixing) the mass
of $h$ beyond 140~GeV (see the next Chapter). A light dominantly
singlet-like state can trigger dominant Higgs-to-Higgs decays of $h$.
\item CP-odd Higgs bosons: The additional dominantly singlet-like state
$A_1$ can again be quite light, triggering Higgs-to-Higgs decays $h \to
A_1 A_1$. Depending on $M_h$ and notably on $M_{A_1}$, many different
cascade decays of $h$ are possible, all of which require dedicated
studies. \end{itemize}

Before we review such studies in Chapter~4, we consider NMSSM specific
phenomena in standard Higgs search channels in the next Chapter.

\section{The NMSSM in Standard Higgs Search Channels}

The establishment of a no-lose theorem in the absence of (dominant)
Higgs-to-Higgs decays in the NMSSM \cite{Ellwanger:2001iw,
Ellwanger:2005uu} relied essentially on the following Higgs production
and decay channels at the LHC (where $H$ denotes any of the three
CP-even Higgs states; see also \cite{Rottlander:2009zz}):

\noi
-- Vector Boson Fusion (VBF) with $H \to \tau^+\tau^-$;\nl
\noi 
-- associate production of $H$ with $W$ or $t\bar{t}$, with $H
\to \gamma\gamma$ and a charged lepton from $W$\nl
\phantom{--} or $t\bar{t}$ in the final state;\nl
\noi
-- associate production of $H$ with $t\bar{t}$, and $H \to b\bar{b}$.

Of course, many more channels contribute to SM-like Higgs searches, as
gluon-gluon ($gg$) fusion and VBF with
$H \to \gamma\gamma$, $H \to W W^{(*)}$, $H \to Z Z^{(*)}$ and various
final states from $W W^{(*)}$, $Z Z^{(*)}$.

The most difficult scenarios in the NMSSM require up to 300 fb$^{-1}$
integrated luminosity at the LHC for a clean signal. These correspond to
cases where the mixing angles in the CP-even Higgs sector are large: The
three physical Higgs states share their couplings to electroweak gauge
bosons according to the sum rule
\beq\label{eq:11}
\sum_{i=1}^3 \bar{g}_i^2 = 1\; ,
\eeq
where $\bar{g}_i$ is the reduced coupling of $H_i$ to $W^\pm$ or $Z$
normalized with respect to the SM. In difficult scenarios, all
$\bar{g}_i$ satisfy $\bar{g}_i^2 \lsim 0.5$. Note that large mixing
angles imply that the mass differences between the CP-even Higgs states
are not large, hence one finds typically $m_{H_i} \lsim 200$~GeV,
$i=1,2,3$, in such scenarios. (Similar observations have been made in
case studies in  \cite{Miller:2004uh,Djouadi:2008uw,Dermisek:2008uu}.)

On the other hand, precisely such scenarios allow potentially for the
simultaneous observation of several Higgs states in the NMSSM, with
masses and couplings incompatible with the MSSM. Corresponding studies
of signal rates for $H_i \to \gamma\gamma$ (production cross sections
times branching ratios) have been performed in \cite{Moretti:2006sv}.
More complete studies (including more relevant Higgs production and
decay channels) concerning the question under which circumstances the
simultaneous observation of several Higgs states would allow to
distinguish the NMSSM from the MSSM would certainly be challenging, but
highly welcome.

In the case of Higgs decays into two photons, already the observation of
a single state can give us possibly a hint in this direction: If the
SM-like and singlet-like states are strongly mixed, the coupling of the
lighter eigenstate to $b$-quarks can be strongly suppressed, implying a
strong reduction of the corresponding partial width into $b \bar{b}$ and
a corresponding enhanced branching ratio into $\gamma\gamma$
\cite{Ellwanger:2010nf,Cao:2011pg}. Inspite the somewhat reduced Higgs
production rate, the signal rate for this process can be six times
larger than in the SM or in the MSSM -- and this for a Higgs mass
possibly well below 114~GeV, but still compatible with LEP constraints
due to the reduced Higgs coupling to the $Z$ boson.

Another feature allowing to distinguish the NMSSM from the MSSM could be
the mass of the mostly SM-like Higgs boson $h$. If this state is the
lightest among all NMSSM CP-even Higgs bosons, the upper bound on its
mass is about $\sim 140$~GeV \cite{Ellwanger:2006rm} for $\l \sim 0.7$
(at the boundary of validity of perturbation theory below the GUT
scale), low $\tan\beta$ and $\k$ such that the negative term
$\Delta_\mathrm{mix}$ in (\ref{eq:4}) is small. However, the mostly
SM-like Higgs boson can well be the next-to-lightest CP-even state $H_2$
in the NMSSM (see Chapter~2), in which case its mass can be larger 
\cite{Ellwanger:1999ji} due to a positive term $\Delta_\mathrm{mix}$ in
(\ref{eq:4}). Then its reduced coupling $\xi \equiv \bar{g}$ to
electroweak gauge bosons is necessarily smaller than~1, in fact $\xi$
decreases with increasing $M_{H_2}$. 

In Table~1 we show three examples of this behaviour, corresponding to
$M_h \equiv M_{H_2} = 140~\mathrm{GeV},\ 145~\mathrm{GeV}$ and
150~GeV\footnote{These results have been obtained with the help of the
code NMHDECAY inside NMSSMTOOLS \cite{Ellwanger:2004xm,
Ellwanger:2005dv}, including the full 1-loop and full $\alpha_s/h_{top}$
two-loop corrections as in \cite{Degrassi:2009yq}. The soft Susy
breaking parameters not shown in Table~1 are 1.2~TeV for the gluino
mass, 1.5~TeV for all squark masses, $A_{top} = -3$~TeV and
$m_{top}=173.1$~GeV.}: $\xi_2$ decreases from 0.92 to 0.73; the lighter
state $H_1$ has $\xi_1 < \xi_2$ allowing it to escape LEP constraints
inspite of its mass down to 91~GeV. Note that $\xi_1^2 + \xi_2^2 \sim
1$; the third CP-even Higgs boson with a mass of about 950~GeV has
$\xi_3 \sim 0$ for the points shown in Table~1. The value $\xi_1 = 0.51$
for $M_{H_1} \sim 97$~GeV seems large at first sight; however, precisely
for this mass range the LEP bounds are particularly weak
\cite{Schael:2006cr} and allow for $\xi^2 \sim 0.25$. For completeness
we also show in Table~1 the branching ratios of the more visible state
$H_2$. The state $H_1$ would be extremely difficult to observe at the
LHC as it decays nearly exclusively into $b\bar{b}$; its branching
ratios into $\gamma\gamma$ are $\lsim 1 \times 10^{-3}$, and into
$\tau^+\tau^-$ about 0.09.

\begin{table}
\begin{center}
\begin{tabular}{|c|c|c|c|} \hline
$\l$ & 0.70 & 0.70  & 0.71 \\\hline
$\k$ & 0.20 & 0.16 & 0.23 \\\hline
$\tan\beta$ & 2.73 & 2.70 & 2.65 \\\hline
$A_\lambda$ & 915~GeV & 928~GeV & 895~GeV \\\hline
$A_\kappa$ & -340~GeV & -230~GeV & -330~GeV \\\hline
$\mu_\mathrm{eff}$ & 320~GeV & 310~GeV & 330~GeV 
\\\hline\hline
$M_{H_2}$ & 140~GeV & 145~GeV  & 150~GeV\\\hline
$\xi_2$ & 0.92 & 0.86 & 0.73 \\\hline
$BR(H_2\to WW)$ & 0.54 & 0.64 & 0.75 \\\hline
$BR(H_2\to ZZ)$ & 0.071 & 0.082 & 0.087 \\\hline
$BR(H_2\to bb)$ & 0.29 & 0.20 & 0.11 \\\hline
$BR(H_2\to \tau\tau)$ & 0.031 & 0.022 & 0.011 \\\hline
$BR(H_2\to \gamma\gamma)$ & $2.3\times 10^{-3}$ & $2.1\times 10^{-3}$ &
$1.7\times 10^{-3}$ \\\hline
$BR(H_2\to Z\gamma)$ & $2.8\times 10^{-3}$ & $2.9\times 10^{-3}$ &
$2.6\times 10^{-3}$ \\\hline
$BR(H_2\to gg,cc)$ & 0.063 & 0.052 & 0.037
\\\hline\hline
$M_{H_1}$ & 91~GeV & 97~GeV  & 115~GeV\\\hline
$\xi_1$ & 0.40 & 0.51 & 0.68 \\\hline
\end{tabular}
\end{center}
\caption{Three examples of scenarios where the state $H_2$ corresponds
to the most SM-like Higgs boson $h$ with a mass $\geq 140$~GeV. We show
its reduced couplings $\xi_2$ to electroweak gauge bosons, its branching
fractions, the mass $M_{H_1}$ of the lighter more singlet-like state as
well as its reduced coupling $\xi_1$.}
\end{table}

To conclude, various NMSSM-specific signals are possible in standard
Higgs search channels at the LHC: signals in the
$W^+W^-/ZZ/\gamma\gamma/b\bar{b}$ final state for Higgs masses $\geq
140$~GeV (incompatible with the MSSM); more visible states than in the
MSSM (although corresponding studies should be extended) and
exceptionally large signal rates in the $\gamma\gamma$ final state,
possibly for unexpectedly small Higgs masses.

\section{Searches for Higgs-to-Higgs Decays}

We have seen in Chapter~2 that many different final states are possible
in the presence of Higgs-to-Higgs decays. Concentrating on $h \to A_1
A_1$, $A_1$ would decay dominantly into $b\bar{b}$ for $M_{A_1} \gsim
10.5$~GeV, into $gg$ for $10.5\ \mathrm{GeV} \gsim M_{A_1} \gsim
9.2$~GeV, into $\tau^+\tau^-$ for $9.2\ \mathrm{GeV} \gsim M_{A_1} \gsim
3.5$~GeV, and into $\mu^+\mu^-$ for $M_{A_1} \lsim 3.5$~GeV. However,
subdominant $A_1$ decays can often lead to more promising signals. With
the exception of gluonic decays due to $A_1 - \eta_b$ mixing, the
subsequent discussion also covers light CP-even states $H_1$ and $h \to
H_1 H_1$ decays (if $h \equiv H_2$). In the present Chapter we review
existing studies on Higgs-to-Higgs decays for the LHC. (An overview over
possible reduced couplings of Higgs bosons, Higgs production cross
sections and branching ratios in various channels and various regions in
the parameter space of the NMSSM is given in \cite{Mahmoudi:2010xp}.)

The first attempt for $M_{A_1} \gsim 10.5$~GeV was made in
\cite{Ellwanger:2003jt} concentrating on $h$ production via Vector Boson
Fusion, where forward and backward jet tagging can be exploited. Not
enforcing b-tagging, the QCD background to the $4b$ final state would
be overwhelming (as for $h$ production via gluon fusion
\cite{Stelzer:2006sp}); hence, the subdominant final state
$2b+2\tau$ was considered. Assuming a value for $M_{A_1}$, two
central jets with $M_{jj} \sim M_{A_1}$ were required. From the two
leptons with the highest transverse momentum and $p_T^\mathrm{miss}$ an
invariant mass $M_{\tau\tau}$ was deduced, and finally the invariant
mass $M_{jj\tau\tau}$ was plotted. A large background comes from
$t\bar{t}$ production. For $L = 300$~fb$^{-1}$, sizeable ratios
$S/\sqrt{B}$ were obtained depending, however, on the accuracy with
which the background shape could be predicted. Moreover, an analysis
including detector simulation and, notably, more realistic lepton
identification efficiency lead to much less optimistic results 
\cite{stefanie}.

Subsequently it was pointed out in \cite{Moretti:2006hq} that
Higgs-Strahlung off $W$ bosons (and, more marginally, off $t\bar{t}$
pairs) can help to establish a signal for $h \to A_1 A_1$ decays, since
one can trigger on an isolated lepton (with, e.g., $p_T \gsim 20$~GeV)
from leptonic $W$ decays. However, only a preliminary analysis of
production cross sections times branching ratios -- without cuts and
background studies -- was performed in \cite{Moretti:2006hq}.

More detailed studies of Higgs-Strahlung off $W$ bosons including
backgrounds, cuts and simulations were performed in
\cite{Cheung:2007sva,Carena:2007jk}. In contrast to
\cite{Ellwanger:2003jt}, $b$-tagging efficiencies of $0.7$ were assumed
in \cite{Cheung:2007sva}, and of $0.5$ (for $E_T^{jet} > 15$~GeV) in
\cite{Carena:2007jk}. This allows to consider the $4b$ final state from
$h \to A_1 A_1 \to 4b$ (and the $2b + 2\tau$ final state
\cite{Carena:2007jk}). Plotting the invariant mass $M_{4b}$, sizeable
significances $S/\sqrt{B} > 5$ for an integrated luminosity $L =
30$~fb$^{-1}$ were found for benchmark points from \cite{Djouadi:2008uw}
with $M_h \sim 110$~GeV and $M_{A_1} \sim 30-40$~GeV in
\cite{Cheung:2007sva}, and for $M_{A_1} \sim \left(M_h -
10\,\mathrm{GeV}\right)/2$ in \cite{Carena:2007jk}. Of course, the
assumed $b$-tagging efficiencies and mistag probabilities are crucial
for these results; hence, corresponding studies including detector
acceptances would be welcome.

In some particular cases, other final states could allow to detect
Higgs-to-Higgs decays: If the branching ratio for $A_1 \to
\gamma\gamma$ is enhanced ($BR(h \to A_1 A_1 \to 4\gamma) \gsim
10^{-4}$), the $4 \gamma$ final state can be visible
\cite{Chang:2006bw}. If $\l$ is very large ($\l \sim 2$ in $\l$Susy
\cite{Cavicchia:2007dp}), $h \equiv H_1$ with a mass $M_h \sim 250$~GeV
will decay dominantly into electroweak gauge bosons leading to
interesting signals in $H_2 \to 2h \to 4(Z\ \mathrm{or}\ W)$ or $A_2 \to
Zh \to ZZZ\ \mathrm{or}\ ZWW$, where $M_{H_2} \sim M_{A_2} \gsim
500$~GeV \cite{Cavicchia:2007dp}.

The $4\tau$ final state will be relevant for small $A_1$ masses. In
\cite{Belyaev:2008gj}, both Higgs-Strahlung and VBF $h$ production
processes were considered, and the $2\,\mu + 2\, j + E_T^{miss}$ final
state from 4~$\tau$-leptons was exploited. After simulation of the
processes, selection cuts were applied and signal cross sections (after
selection cuts) were given for a range of parameters corresponding to
$M_{A_1} < 10$~GeV, $M_h$ from 20 to 130~GeV (respecting LEP constraints
\cite{Schael:2006cr} before the ALEPH analysis \cite{Schael:2010aw}).
Notably for $M_h \gsim 100$~GeV (hardly affected by ALEPH constraints
\cite{Schael:2010aw}), signal cross sections up to 10~fb
(Higgs-Strahlung) and 80~fb (VBF) were found. However, backgrounds and
detector performances have not been included in this study.

An extensive study on searches for $h \to A_1 A_1 \to 4\tau$, including
all $h$ production processes, background processes and the performances
of the ATLAS detector, was performed in \cite{Rottlander:2008zz}. For
$M_h$ from 100--130~GeV and $5\ \mathrm{GeV} < M_{A_1} < 10$~GeV, a
signal significance of $\sim 5$ was obtained for $L = 30$~fb$^{-1}$ in
the $4\mu + E_T^{miss}$ final state from $h \to A_1 A_1 \to 4\tau \to
4\mu\ +$ neutrinos in $h$ production via VBF.

The subdominant decay channel $h \to A_1 A_1 \to 2\mu+ 2\tau+
E_T^{miss}$ (with $h$ from gluon fusion) was analysed in 
\cite{Lisanti:2009uy}. Inspite of the reduction of the signal rate by
the factor $(m_\mu/m_\tau)^2$, it was argued that the clean signal in
the dimuon invariant mass allows to cover most of the relevant region of
the parameter space already with $L \sim 5$~fb$^{-1}$ at the LHC.
(However, the magnitude of the QCD multijet background estimated in
\cite{Lisanti:2009uy} was found to be three orders of magnitude larger
in \cite{Belyaev:2010ka}.)

Another possibility for the study of $h \to A_1 A_1 \to 4\tau$ is the
central exclusive production (CEP) of $h$, $pp \to p+h+p$
\cite{Forshaw:2007ra}. This requires the installation of forward proton
detectors in the high dispersion region as in the FP420 project
\cite{Albrow:2008pn}. According to the results of simulations of the
signal and backgrounds including pile up in \cite{Forshaw:2007ra}, 
a significant signal can be obtained for sufficient instantaneous and
integrated luninosity.

If $M_{A_1} < 2\,m_\tau$, the process $h \to A_1 A_1 \to 4\mu$ is very
promising. Analyses of the $h$ production cross sections (via gluon
fusion and associate production with $b\bar{b}$) times branching
fractions have been performed in \cite{Belyaev:2010ka}, and were
compared to the QCD multijet background. Requiring at least one $\mu$
with $p_T > 20$~GeV (4 muons with $p_T > 5$~GeV) and plotting invariant
masses of opposite charge dimuon pairs as well as the $M_{4\mu}$
invariant mass, most of the parameter space corresponding to $M_{A_1} <
2\,m_\tau$ allows for the detection of both $A_1$ and $h$ already for $L
\sim 1$~fb$^{-1}$ at the LHC \cite{Belyaev:2010ka}.

In the case of large $\tan\beta$, the associate production of $h$ with
$b$-quarks is interesting. In \cite{Almarashi:2011te}, signal
rates for $pp \to b\bar{b}h \to b\bar{b}A_1A_1$ (and $pp \to b\bar{b}h_2
\to b\bar{b}h_1h_1$) times branching fractions into $4\gamma$, $4b$,
$2b\,2\tau$, $4\tau$ and $4\mu$ final states (in addition to the prompt
$b$-quark pair) are given, but more dedicated analyses are required in
these cases.

Another potentially interesting Higgs-to-Higgs decay process is the
decay of a charged Higgs boson $H^\pm$ into $W^\pm + A_1$ (or $W^\pm +
h$ \cite{Drees:1999sb}). Branching ratios for $H^\pm \to W^\pm + A_1/h$
and cross sections for the processes $pp \to H^\pm A_1 \to W^\pm A_1
A_1$ (and $pp \to W^\pm h \to W^\pm A_1 A_1$, which can be of similar
order) are given in \cite{Akeroyd:2007yj}, the branching ratios for
$H^\pm \to W^\pm + A_1$ have also been studied in
\cite{Dermisek:2008uu,Mahmoudi:2010xp}.

A light NMSSM specific CP-odd Higgs boson $A_1$ might also be visible in
direct production channels, without relying on $h \to A_1 A_1$ decays.
For $M_{A_1} \lsim 12$~GeV, the (subdominant) decay $A_1 \to
\mu^+ \mu^-$ can allow for $A_1$ detection via gluon-gluon fusion
\cite{Dermisek:2009fd} due to the clean signal (for sufficiently large
$\tan\beta$, such that the $b$-quark-loop induced production rate is
sufficiently large in spite of the dominantly singlet-like nature of
$A_1$). First searches by ATLAS based on 35.4~pb$^{-1}$ integrated
luminosity did not discover a signal \cite{ATLAS-CONF-2011-020}, but
with more accumulated data the prospects will be more promising.

At large $\tan\beta$, the associate production of $A_1$ with a $b
\bar{b}$ pair can become relevant for $M_{A_1}$ up to $M_Z$
\cite{Almarashi:2010jm,Almarashi:2011hj,Almarashi:2011bf}. The
two-photon and $\tau^+\tau^-$ decay modes of $A_1$ have been analysed in
\cite{Almarashi:2010jm}, where appropriate cuts have been applied and
signal-to-backbround ratios been studied for 300~fb$^{-1}$ integrated
luminosity at the LHC. The two-photon decay mode of $A_1$ seems too
small, but the $\tau^+\tau^-$ decay mode can lead to a sufficiently
large signal-to-backbround ratio. The subdominant $\mu^+\mu^-$ decay
mode of $A_1$ has been analysed in \cite{Almarashi:2011hj}. It can lead
to a signal for 30~fb$^{-1}$ integrated luminosity if $M_{A_1}$ is in
the range $10 - 40$~GeV, whereas more integrated luminosity would be
required for larger values of $M_{A_1}$. In the $4b$ final state and
with high $b$-tagging efficiency, a signal may be visible for $M_{A_1}$
in the range $20 - 80$~GeV \cite{Almarashi:2011bf}.

Light Higgs bosons of the NMSSM could also be produced in sparticle
decay cascades. Branching fractions for neutralino decays into
neutralinos plus $A_1$ have been studied in
\cite{Choi:2004zx,Cheung:2008rh}, and for sbottom/stau decays into
sbottom/stau plus $A_1$ in \cite{Kraml:2005nx}. Simulations of such
processes have not been performed, with the exception of gluonic decays
of $A_1$ (see below) in \cite{Bellazzini:2010uk}.

Clearly a dominant $A_1$ decay into two gluons, and hence a dominant $h$
decay into $h \to A_1 A_1 \to 4g$, would constitute a major challenge
for $h$ detection at the LHC. As discussed in Chapter~2 this would
happen for $9.2\ \mathrm{GeV} \lsim M_{A_1} \lsim 10.5$~GeV, in which case
the search modes discussed above would fail. Recently it has been
proposed that the analysis of jet substructures could come to the rescue
in such situations \cite{Chen:2010wk,Falkowski:2010hi,
Bellazzini:2010uk, Kaplan:2011vf}.

Here one concentrates on $h$ production in association with
a $W$ boson, where an isolated lepton from the $W$ decay helps to
trigger on the events \cite{Chen:2010wk,Falkowski:2010hi,Kaplan:2011vf}.
($h$ production in sparticle decay cascades has been considered in this
context in \cite{Bellazzini:2010uk}.) In addition one requires two jets
with large $p_T$, which originate from decays $h \to A_1 A_1 \to 2j$
with boosted $A_1$ bosons. Decays of boosted Higgs bosons allow to
search for jet substructures \cite{Butterworth:2008iy,
Abdesselam:2010pt}. Here one assumes that the decay $A_1 \to 2g$ gives
rise to a single ``fat'' jet $j$, whose substructure can be analysed:
undoing the last recombination step of the clustering algorithm which
generated the jet $j$ leads to the decomposition $j \to \{j_1,\,j_2\}$.
Typically one requires $m_{j_1} \sim m_{j_2} \ll m_j \lsim 12$~GeV, and
not more jets with large $p_T$ than required for a signal. Plotting
$m_{jj}$ of the events satisfying corresponding criteria
\cite{Chen:2010wk,Falkowski:2010hi, Bellazzini:2010uk, Kaplan:2011vf}
can lead to visible peaks for $m_{jj} \sim m_h$.

Analyses based on jet substructure are not confined to dominant gluonic
decays of $A_1$; they can also be applied to larger $A_1$ masses leading
to dominant $A_1 \to b\bar{b}$ decays \cite{Kaplan:2011vf} and, notably,
to $A_1 \to 2\tau$ decays \cite{Englert:2011iz} where two hadronically
decaying $\tau$ leptons from a boosted $A_1$ form a single ``fat'' ditau
jet.

For convenience we have summarized the available studies of
Higgs-to-Higgs decays and single $A_1$ production processes at the LHC,
for different ranges of $M_{A_1}$ and ordered according to production
processes and final states, in Table~2. Some of these studies are
confined to estimates of production cross sections times branching
fractions. Studies including simulations of background processes and
estimates of signal to background ratios after cuts are indicated by an
asterisk; two asterisks indicate studies including detector simulations.
(The gluonic decay $A_1 \to gg$ is left aside in Table~2.) 

It should also be noted that in many cases specific ranges of parameters
are required, such that the production cross sections and/or branching
fractions are large enough allowing for sufficiently significant
signals. Hence the existence of a study of a given channel implies by no
means that a discovery of the corresponding process (for corresponding
Higgs masses) is guaranteed; moreover, simulations including the
detector response are missing in most cases.

\begin{table}
\begin{center}
\begin{tabular}{|p{27mm}|c|c|c|c|c|c|c|} \hline
$M_{A_1}$:& \multicolumn{4}{c|} {$\gsim 10.5$~GeV}
& \multicolumn{2}{c|} {$\lsim 10.5$~GeV}
& $ < 2\,m_\tau$ 
\\\hhline{|========|}
& \multicolumn{7}{c|} {$h \to A_1\,A_1$}\\\hline
Final state: & $4b$ &\multicolumn{2}{c|} {$2b+2\tau$} & $4\gamma$ & 
$4\tau$ & $2\tau+2\mu$ & $4\mu$ \\\hline
$h$ production\nl mode:& &\multicolumn{2}{c|}{} & & & & \\\hline
VBF: & &
\multicolumn{2}{c|}{\cite{Ellwanger:2003jt}$^*$, \cite{stefanie}$^{**}$}
 & &\cite{Belyaev:2008gj}$^*$, \cite{Rottlander:2008zz}$^{**}$ 
 & & \\\hline
 $W/Z+h$: &\cite{Cheung:2007sva}$^*$, \cite{Carena:2007jk}$^*$ 
 &
 \multicolumn{2}{c|}{\cite{Moretti:2006hq}} & &\cite{Belyaev:2008gj}$^*$,
 \cite{Englert:2011iz}$^*$ & & \\\hline
$gg$: & &\multicolumn{2}{c|}{} &\cite{Chang:2006bw}$^*$ & &
\cite{Lisanti:2009uy}$^*$ &\cite{Belyaev:2010ka}$^*$  \\\hline
CEP: & &\multicolumn{2}{c|}{} & &\cite{Forshaw:2007ra}$^{**}$ & &
\\\hline
$b\overline{b}h$:
&\cite{Almarashi:2011te} &
\multicolumn{2}{c|}{\cite{Almarashi:2011te}} &
\cite{Almarashi:2011te} &\cite{Almarashi:2011te} & &
\cite{Belyaev:2010ka}$^*$  \\\hhline{|========|}
& \multicolumn{7}{c|} {Single $A_1$ production}\\\hline
Final state:& $2b$ & $2\tau$ & $2\mu$ & $2\gamma$ & $2\tau$ & {$2\mu$}
& {$2\mu$}\\\hline
$A_1$ production\nl mode:& & & & & & & \\\hline
$gg$: & & & & & &\cite{Dermisek:2009fd}$^*$ & \\\hline
$b\overline{b}A_1$: 
&\cite{Almarashi:2011bf}$^*$ &\cite{Almarashi:2010jm}$^*$&
\cite{Almarashi:2011hj}$^*$  &\cite{Almarashi:2010jm}$^*$ &
\cite{Almarashi:2010jm}$^*$ &\cite{Almarashi:2011hj}$^*$  &
\cite{Almarashi:2011hj}$^*$  \\\hline
$H^\pm \to W^\pm A_1$&\cite{Akeroyd:2007yj} & & & &
\cite{Akeroyd:2007yj} & & \\\hline
\end{tabular}
\end{center}
\caption{Available studies of Higgs-to-Higgs decays and single $A_1$
production at the LHC, for different ranges of $M_{A_1}$. (CEP stands
for central exclusive production.) []$^*$ indicates that, apart from
signal rates, signal and background processes have been simulated or
estimated. []$^{**}$ indicates that, in addition, the detector response
has been included in the study. (The gluon-gluon final state, relevant
for $M_{A_1}$ in the range $M_{A_1} \sim 10\pm 0.5$~GeV, has been left
aside.)}
\end{table}

\section{Conclusions}

In various regions of the parameter space of the NMSSM, the properties
of Higgs bosons are clearly distinct from the MSSM: A completely SM-like
Higgs boson can have a mass up to 140~GeV, and a dominantly SM-like
Higgs boson (with somewhat reduced couplings to electroweak gauge
bosons) can be heavier. In this case this Higgs boson has a
non-vanishing singlet component, and is necessarily accompagnied by a
lighter state which is equally a doublet-singlet admixture with a mass
and couplings typically allowed by LEP constraints.

In other regions of the parameter space, branching ratios into two
photons can be enhanced due to a strong suppression of the partial width
into $b\bar{b}$.

In spite of numerous studies of Higgs-to-Higgs decays, a no-lose theorem
could not be established up to now. For $M_{A_1} \gsim 10.5$~GeV, the
most promising studies in \cite{Cheung:2007sva,Carena:2007jk} on the
$4b$ final state must be confirmed with respect to the assumed
$b$-tagging efficiencies, mistaggings and backgrounds by studies
including detector simulations. The latter are also required for the
analyses of the gluonic decays for $M_{A_1}$ around 10~GeV in
\cite{Chen:2010wk,Falkowski:2010hi, Bellazzini:2010uk, Kaplan:2011vf},
where the study of jet substructures requires measurements of invariant
masses of very slim (but very boosted) jets. For $M_{A_1} \lsim 9.2$~GeV
and dominant $h \to A_1 A_1 \to 4\tau$ decays, the studies in
\cite{Rottlander:2008zz,Forshaw:2007ra} seem promising, whereas the QCD
background assumed in the analysis of the $2\tau 2\mu$ final state in
\cite{Lisanti:2009uy} needs to be confirmed. Background and detector
simulations are also required for the proposals for direct $A_1$
production in \cite{Dermisek:2009fd,Almarashi:2010jm,Almarashi:2011hj,
Almarashi:2011bf}, where the significance of the signal rates depend on
model parameters as $\tan\beta$. Hence, further studies on
Higgs-to-Higgs decays are still necessary, if one wants to be sure that
at least one Higgs boson of the NMSSM is visible at the LHC.

\section*{Acknowledgements}

The author acknowledges support from the French ANR LFV-CPV-LHC.

\end{document}